\documentstyle[epsf,12pt]{article}

\begin{document}

\title
{Anomaly puzzle in $N=1$ supersymmetric electrodynamics as
artifact of dimensional reduction.}

\author{K.V.Stepanyantz \thanks{E-mail:$stepan@theor.phys.msu.su$}}

\maketitle

\begin{center}

{\em Moscow State University, physical faculty,\\
department of theoretical physics.\\$117234$, Moscow, Russia}

\end{center}

\begin{abstract}
Calculations of the two-loop $\beta$-function for $N=1$ supersymmetric
electrodynamics are compared for regularizations by higher derivatives
and by the dimensional reduction. The renormalized effective action are
found to be the same for both regularizations. However, unlike the
dimensional reduction, the higher derivative regularization does not
lead to anomaly puzzle, because it allows to perform correct calculation
of diagrams with insertions of counterterms. In particular, using this
method a contribution of diagrams with insertions of counterterms is
calculated exactly to all orders. This contribution appears to be 0 if
the theory is regularized by the dimensional reduction. We argue, that
this result follows from mathematical inconsistency of the dimensional
reduction and is responsible for the anomaly puzzle.
\end{abstract}

\sloppy


\section{Introduction.}
\hspace{\parindent}

It is well known \cite{Ferrara,Clark,Piquet1,Piquet2}, that in
supersymmetric theories the axial and the trace of the energy-momentum
tensor anomalies are components of a chiral scalar supermultiplet.
Adler-Bardeen theorem \cite{Bardeen,Slavnov_Book} asserts that there
are no radiative corrections to the axial anomaly beyond the one-loop
approximation, while the trace anomaly is proportional to the
$\beta$-function \cite{Adler_Collins} to all orders. Therefore this
seems to imply, that the $\beta$-function in supersymmetric theories
should be exhausted by the first loop \cite{NSVZ_PL}. It really takes
place in models with $N=2$ supersymmetry \cite{N2}. However explicit
perturbative calculations find higher order corrections to the
$\beta$-functions of $N=1$ supersymmetric theories, regularized by
dimensional reduction \cite{Tarasov,Grisaru,Caswell}. This contradiction
is usually called "anomaly puzzle".

Many papers are devoted to attempts of solving the anomaly puzzle in
supersymmetric theories. For example, in \cite{SV} the anomaly puzzle
is argued to be a consequence of the difference between usual and
Wilsonian effective actions. In particular, the authors noted, that the
nontrivial contribution to the $\beta$-function come from the so-called
Konishi anomaly \cite{Konishi,ClarkKonishi}. Investigation of this
contribution in \cite{SV} and investigation of instanton contributions
in \cite{NSVZ_Instanton} lead to construction of the so called exact
Novikov, Shifman, Vainshtein and Zakharov (NSVZ) $\beta$-function.
For $N=1$ supersymmetric electrodynamics considered in this paper
the NSVZ $\beta$-function has the following form:

\begin{equation}\label{NSVZ_Beta}
\beta(\alpha) = \frac{\alpha^2}{\pi}\Big(1-\gamma(\alpha)\Big)
\end{equation}

\noindent
where $\gamma(\alpha)$ is the anomalous dimension of the matter superfield.
Explicit perturbative calculations using regularization by the dimensional
reduction (DRED) verify the NVSZ $\beta$-function up to two-loop order.
Nevertheless, three loop results obtained in
\cite{ThreeLoop1,ThreeLoop2,ThreeLoop3} do not agree with the NSVZ
$\beta$-function. However \cite{Jones} this disagreement can be eliminated
by a special choice of renormalization scheme, a possibility of such
choice being highly nontrivial \cite{JackJones}. In principle it is
possible to relate $\overline{\mbox{DRED}}$ scheme and NSVZ scheme order
by order \cite{North} in the perturbation theory.

A very simple and beautiful solution of anomaly puzzle, different from the
solution of \cite{SV}, was presented in \cite{Arkani}. The main idea of
this paper is that the higher order corrections in NSVZ $\beta$-function
are due to anomalous Jacobian under the rescaling of the fields done in
passing from holomorphic to canonical normalization. In the case of
supersymmetric electrodynamics holomorphic normalization means, that the
renormalized action is written as

\begin{eqnarray}
&& S_{ren} =
\frac{1}{4 e^2} Z_3(\Lambda/\mu) \mbox{Re}
\int d^4x\,d^2\theta\,W_a C^{ab} W_b
+\nonumber\\
&& \qquad\qquad\qquad\qquad\qquad
+ Z(\Lambda/\mu)\frac{1}{4}\int d^4x\, d^4\theta\,
\Big(\phi^* e^{2V}\phi +\tilde\phi^* e^{-2V}\tilde\phi\Big),\qquad
\end{eqnarray}

\noindent
while in the canonical normalization

\begin{eqnarray}
&& S_{ren} =
\frac{1}{4 e^2} Z_3(\Lambda/\mu) \mbox{Re}
\int d^4x\,d^2\theta\,W_a C^{ab} W_b
+\nonumber\\
&& \qquad\qquad\qquad\qquad\qquad\qquad\quad
+ \frac{1}{4}\int d^4x\, d^4\theta\,
\Big(\phi^* e^{2V}\phi +\tilde\phi^* e^{-2V}\tilde\phi\Big),\qquad
\end{eqnarray}

\noindent
In the former case the $\beta$-function is supposed to be exhausted at
the one loop, while in the latter case it coincides with the NVSZ result.
In principle this solution is different from the one, given by Shifman
and Vainshtein. Moreover, it contradicts to the results of explicit
two-loop calculations, made by dimensional reduction. The authors of
\cite{Arkani} supposed, that in the holomorphic normalization the
$\beta$-function is exhausted at the one loop if one uses higher
covariant derivative regularization \cite{Slavnov,Bakeyev}, supplemented
by the Pauli-Villars. It is known \cite{PhysLett}, that this regularization
always yields the same result for one-loop logarithmic divergences as the
dimensional regularization (reduction). Explicit two-loop calculations
for theories, regularized by higher derivatives (HD), were made first in
\cite{hep,tmf2} for $N=1$ supersymmetric electrodynamics and gave zero
two-loop contribution to the $\beta$-function. This result implies absence
of anomaly puzzle in view of the solution proposed in \cite{Arkani}.
However it is not quite clear why different regularizations give different
results for the scheme independent two-loop $\beta$-function.
In principle, in \cite{hep} we noted, that using of the higher derivative
regularization leads to a nontrivial contribution of diagrams with
insertions of one-loop counterterms, which is absent for dimensional
reduction. However the detailed analysis of this result was not yet made.

In this paper calculation of diagrams with insertions of counterterms
is analyzed for the DRED and HD regularization. DRED technique proposed
by Siegel \cite{Siegel} consists of continuing in the number of space-time
dimensions from 4 to $n$, where $n$ is less than 4, but keeping the
numbers of components of all other tensors fixed. It is important to note,
that such regularization is mathematically inconsistent \cite{Siegel2}.
As a consequence, a straightforward application of DRED to the calculation
of axial anomaly gives incorrect zero result, because DRED does not break
chiral symmetry. It is necessary to stress an essential difference between
the DRED and the dimensional regularization (DREG), which is mathematically
consistent and allows to calculate anomalies \cite{tHV}. In principle,
it is possible to calculate axial anomaly even within the DRED technique.
However, for this purpose it is necessary to impose some mathematically
inconsistent conditions, for example $\mbox{tr}(AB)\ne \mbox{tr}(BA)$
\cite{Nikolai}. Another possibility is an attempt to go to $n>4$ in the
DRED scheme \cite{Leveille}, that also leads to some contradictions.

In the present paper we argue, that the above mentioned contradictions
of DRED lead to the incorrect result for the sum of diagrams with
insertions of counterterms and, therefore, for the $\beta$-function. These
arguments are confirmed by comparison between calculations of the two-loop
$\beta$-function for the $N=1$ supersymmetric electrodynamics by DRED and
HD regularization.

The paper is organized as follows:

In section \ref{Section_SUSY_QED} we consider $N=1$ supersymmetric
electrodynamics and different ways of its regularization. In particular,
in this section we remind main contradictions of DRED, pointed out by
Siegel. Calculation of the two-loop contribution to the
$\beta$-function and relationship between this contribution and Konishi
anomaly are analyzed in section \ref{Section_Two_Loop} using different
regularizations. In section \ref{Section_Exact} the sum of diagrams with
insertions of counterterms on the matter lines regularized by HD is
calculated exactly to all orders. Section \ref{Section_Conclusion}
contains some concluding remarks. Finally, appendix contains a derivation
of the exact result for a sum of diagrams with insertions of counterterms.


\section{Supersymmetric electrodynamics and its regularization.}
\label{Section_SUSY_QED}

\subsection{$N=1$ supersymmetric electrodynamics.}
\hspace{\parindent}

$N=1$ supersymmetric electrodynamics in the superspace is described
by the following action:

\begin{equation}\label{SQED_Action}
S_0 = \frac{1}{4 e^2} \mbox{Re}\int d^4x\,d^2\theta\,W_a C^{ab} W_b
+ \frac{1}{4}\int d^4x\, d^4\theta\,
\Big(\phi^* e^{2V}\phi +\tilde\phi^*
e^{-2V}\tilde\phi\Big).
\end{equation}

\noindent
Here $\phi$ and $\tilde\phi$ are chiral superfields

\begin{eqnarray}\label{Phi_Superfield}
&& \phi(y,\theta) = \varphi(y) + \bar\theta (1+\gamma_5) \psi(y)
+ \frac{1}{2}\bar\theta (1+\gamma_5)\theta f(y);\nonumber\\
&&\tilde \phi(y,\theta) = \tilde \varphi(y)
+ \bar\theta (1+\gamma_5) \tilde \psi(y)
+ \frac{1}{2}\bar\theta (1+\gamma_5)\theta \tilde f(y),
\end{eqnarray}

\noindent
where $y^\mu = x^\mu + i\bar\theta\gamma^\mu\gamma_5\theta/2$.
Two Majorana spinors $\psi$ and $\tilde\psi$ form one Dirac spinor

\begin{equation}\label{Psi_Definition}
\Psi = \frac{1}{\sqrt{2}}\Big((1+\gamma_5)\psi+(1-\gamma_5)\tilde\psi\Big).
\end{equation}

\noindent
$V$ in (\ref{SQED_Action}) is a real superfield

\begin{eqnarray}\label{V_Superfield}
&& V(x,\theta) = C(x)+i\sqrt{2}\bar\theta\gamma_5\xi(x)
+\frac{1}{2}(\bar\theta\theta)K(x)
+\frac{i}{2}(\bar\theta\gamma_5\theta)H(x)
+\frac{1}{2}(\bar\theta \gamma^\mu \gamma_5\theta) A_\mu(x)
+\nonumber\\
&& + \sqrt{2} (\bar\theta\theta) \bar\theta
\Big(i\gamma_5\chi(x)
+\frac{1}{2}\gamma^\mu\gamma_5\partial_\mu\xi(x)\Big)
+ \frac{1}{4} (\bar\theta\theta)^2 \Big(D(x)
-\frac{1}{2}\partial^2 C(x)\Big),
\end{eqnarray}

\noindent
where, in particular, $A_\mu$ is an Abelian gauge field. The superfield
$W_a$ in the Abelian case is defined by

\begin{equation}
W_a = \frac{1}{16} \bar D (1-\gamma_5) D\Big[(1+\gamma_5)D_a V\Big],
\end{equation}

\noindent
where $D$ is the supersymmetric covariant derivative

\begin{equation}
D = \frac{\partial}{\partial\bar\theta} - i\gamma^\mu\theta\,\partial_\mu.
\end{equation}


\subsection{Higher derivative regularization.}
\hspace{\parindent}

In order to regularize model (\ref{SQED_Action}) by HD its action should
be modified as follows:

\begin{eqnarray}\label{Regularized_SQED_Action}
&& S_0 \to S = S_0 + S_{\Lambda}
=\vphantom{\frac{1}{2}}\nonumber\\
&&\qquad
= \frac{1}{4 e^2} \mbox{Re}\int d^4x\,d^2\theta\,W_a C^{ab}
\Big(1+ \frac{\partial^{2n}}{\Lambda^{2n}}\Big) W_b
+\nonumber\\
&&\qquad\qquad\qquad\qquad\qquad\qquad
+ \frac{1}{4}\int d^4x\, d^4\theta\,
\Big(\phi^* e^{2V}\phi +\tilde\phi^* e^{-2V}\tilde\phi\Big).\qquad
\end{eqnarray}

\noindent
Note, that in the Abelian case the superfield $W^a$ is gauge invariant,
so that the higher derivative term contains usual derivatives.

Quantization of (\ref{Regularized_SQED_Action}) can be made using
standard technique described in \cite{West} and is not considered here.
It is necessary to mention only that the gauge invariance was fixed by
adding of

\begin{equation}
S_{gf} = - \frac{1}{64 e^2}\int d^4x\,d^4\theta\,
\Bigg(V D^2 \bar D^2
\Big(1 + \frac{\partial^{2n}}{\Lambda^{2n}}\Big) V
+ V \bar D^2 D^2
\Big(1+ \frac{\partial^{2n}}{\Lambda^{2n}}\Big) V\Bigg),
\end{equation}

\noindent
where

\begin{equation}
D^2 \equiv \frac{1}{2} \bar D (1+\gamma_5)D;\qquad
\bar D^2 \equiv \frac{1}{2}\bar D (1-\gamma_5) D.
\end{equation}

\noindent
After adding of such terms the free part of the action for the
superfield $V$ is written in the most simple form

\begin{equation}
S_{gauge} + S_{gf} = \frac{1}{4 e^2}\int d^4x\,d^4\theta
V\partial^2 \Big(1+ \frac{\partial^{2n}}{\Lambda^{2n}}\Big) V.
\end{equation}

\noindent
In the Abelian case diagrams containing ghost loops are certainly absent.

The superficial degree of divergence for the model
(\ref{Regularized_SQED_Action}) is equal to (see e.f. \cite{hep})

\begin{equation}\label{Degree_Of_Divergence}
\omega_\Lambda = 2 - 2n (L-1) - E_\phi (n+1),
\end{equation}

\noindent
where $L$ is a number of loops and $E_\phi$ is a number of external
$\phi$-lines. According to (\ref{Degree_Of_Divergence}) divergences
remain in one-loop diagrams even for $n\ge 2$. In order to regularize
these divergences it is necessary to insert in the generating functional
Pauli-Villars determinants \cite{Slavnov_Book}:

\begin{eqnarray}\label{Modified_Z}
&& Z = \int DV\,D\phi\,D\tilde \phi\,
\prod\limits_i \Big(\det PV(V,M_i)\Big)^{c_i}
\exp\Bigg\{i\Bigg[\frac{1}{4 e^2} \int d^4x\,d^4\theta\, V\partial^2
\Big(1+ \frac{\partial^{2n}}{\Lambda^{2n}}\Big) V
-\nonumber\\
&& - \frac{1}{4 e^2} \Big(Z_3(\Lambda/\mu)-1\Big) \int d^4x\,d^4\theta\,
V \Pi_{1/2}\partial^2
\Big(1+ \frac{\partial^{2n}}{\Lambda^{2n}}\Big) V
+\nonumber\\
&& \qquad\qquad\qquad\qquad\qquad\qquad\qquad\qquad
+ \frac{1}{4} Z(\Lambda/\mu) \int d^4x\,d^4\theta\,
\Big(\phi^* e^{2V}\phi
+ \tilde\phi^* e^{-2V}\tilde\phi \Big)
+\nonumber\\
&&
+ \int d^4x\,d^4\theta\,J V
+ \int d^4x\,d^2\theta\, \Big(j\,\phi + \tilde j\,\tilde\phi \Big)
+ \int d^4x\,d^2\bar\theta\,
\Big(j^*\phi^* + \tilde j^* \tilde\phi^* \Big)\Bigg]\Bigg\},
\end{eqnarray}

\noindent
where

\begin{eqnarray}\label{PV_Determinants}
&& \Big(\det PV(V,M)\Big)^{-1} = \int D\Phi\,D\tilde \Phi\,
\exp\Bigg\{i\Bigg[ Z(\Lambda/\mu) \frac{1}{4} \int d^4x\,d^4\theta\,
\Big(\Phi^* e^{2V}\Phi
+\qquad\nonumber\\
&& + \tilde\Phi^* e^{-2V}\tilde\Phi \Big)
+ \frac{1}{2}\int d^4x\,d^2\theta\, M \tilde\Phi \Phi
+ \frac{1}{2}\int d^4x\,d^2\bar\theta\, M \tilde\Phi^* \Phi^*
\Bigg]\Bigg\},
\end{eqnarray}

\noindent
and the coefficients $c_i$ satisfy equations

\begin{equation}
\sum\limits_i c_i = 1;\qquad \sum\limits_i c_i M_i^2 = 0.
\end{equation}

\noindent
Below we will assume, that $M_i = a_i\Lambda$, where $a_i$ are some
constants. Insertion of Pauli-Villars determinants allows to cancel
remaining divergences in all one-loop diagrams, including diagrams
with insertions of counterterms.

In our notations the generating functional $W$ is defined by

\begin{equation}\label{W}
W = - i\ln Z
\end{equation}

\noindent
and an effective action is obtained by making a Legendre transformation:

\begin{equation}\label{Gamma}
\Gamma = W - \int d^4x\,d^4\theta\,J V
- \int d^4x\,d^2\theta\, \Big(j\,\phi + \tilde j\,\tilde\phi \Big)
- \int d^4x\,d^2\bar\theta\,
\Big(j^*\phi^* + \tilde j^* \tilde\phi^* \Big),
\end{equation}

\noindent
where $J$, $j$ and $\tilde j$ is to be eliminated in terms of
$V$, $\phi$ and $\tilde\phi$, through solving equations

\begin{equation}
V = \frac{\delta W}{\delta J};\qquad
\phi = \frac{\delta W}{\delta j};\qquad
\tilde\phi = \frac{\delta W}{\delta\tilde j}.
\end{equation}

Due to the supersymmetric gauge invariance

\begin{equation}
V \to V - \frac{1}{2}(A+A^+);
\qquad \phi\to e^{A}\phi;\qquad \tilde\phi\to e^{-A} \tilde\phi,
\end{equation}

\noindent
where $A$ is an arbitrary chiral scalar superfield, the renormalized
action can be written as

\begin{eqnarray}\label{Renormalized_Action}
&& S_{ren} =
\frac{1}{4 e^2} Z_3(\Lambda/\mu) \mbox{Re}\int d^4x\,d^2\theta\,W_a C^{ab}
\Big(1+ \frac{\partial^{2n}}{\Lambda^{2n}}\Big) W_b
+\nonumber\\
&& \qquad\qquad\qquad\qquad\qquad
+ Z(\Lambda/\mu)\frac{1}{4}\int d^4x\, d^4\theta\,
\Big(\phi^* e^{2V}\phi +\tilde\phi^* e^{-2V}\tilde\phi\Big).\qquad
\end{eqnarray}

\noindent
Here $e=e(\Lambda/\mu)$ is a renormalized coupling constant, while
the bare coupling constant $e_0$ is given by

\begin{equation}\label{Bare_Coupling_Constant}
\frac{1}{e_0^2} = \frac{1}{e(\Lambda/\mu)^2} Z_3(\Lambda/\mu)
\end{equation}

\noindent
and does not depend on $\mu$.

Having obtained $S_{ren}$, it is possible to find the $\beta$-function
and the anomalous dimension, which in our notations are defined by

\begin{equation}\label{Beta_Gamma_Definition}
\beta = \frac{d}{d\ln\mu}\Bigg(\frac{e^2}{4\pi}\Bigg);
\qquad\quad
\gamma = \frac{d\ln Z}{d\ln\mu}.
\end{equation}


\subsection{Dimensional reduction and dimensional regularization.}
\hspace{\parindent}

Although HD regularization can be easily applied to calculations of
quantum corrections in the supersymmetric electrodynamics
\cite{hep,tmf2,tmf1}, use of this regularization encounters considerable
technical difficulties in non-Abelian gauge theories due to complicated
structure of vertexes. That is why the HD regularization was not applied
to calculations so often as the DREG \cite{tHV} or DRED \cite{Siegel}.

In the DREG method calculations of quantum corrections are formally
performed in the space-time with dimension $n\ne 4$. However this method
is not well-suited for supersymmetric theories, because DREG does not
preserve invariance of the action with respect to the supersymmetry
transformations. The reason is that a necessary condition for
supersymmetry is equality of Bose and Fermi degrees of freedom, which
can take place only for integer $n$.

A modification of DREG so as to render it compatible with supersymmetry
was made by Siegel \cite{Siegel}. According to Siegel's method the $n=4$
Lagrangian is dimensionally reduced to $n<4$ dimensions. Then a vector
$A_\mu$ is split into an $n$ component vector $\tilde A_\mu$ and
$\epsilon=4-n$ "$\epsilon$-scalars" $\hat A_{\mu}$, but the total number
of bosons remains $n$ independent. It is important, that the dimension $n$
should be less than 4, so that

\begin{equation}
\delta_\mu^\nu \tilde \delta^\mu_\nu = n,
\end{equation}

\noindent
where $\delta_\mu^\nu$ is 4-dimensional Kronecker symbol and
$\tilde \delta_\mu^\nu$ is $n$-dimensional Kronecker symbol.

However as pointed out in \cite{Siegel2} there remain ambiguities with
dimensional reduction associated with treatment of the Levi-Civita symbol.
For example, the product

\begin{equation}
\tilde\varepsilon^{\alpha\beta\gamma\delta}
\tilde \varepsilon_{\alpha\beta\gamma\delta}
\,\hat\varepsilon^{\mu\nu\rho\tau}
\hat \varepsilon_{\mu\nu\rho\tau}
\end{equation}

\noindent
depends on a way of calculation and can be equal either to $0$ or to

\begin{equation}
(n-4)(n-3)^2(n-2)^2(n-1)^2 n,
\end{equation}

\noindent
Therefore the dimension reduction is mathematically consistent only for
integer $n\le 4$. Moreover, if $n<4$ the $\gamma_5$-matrix should be
chosen so that

\begin{equation}\label{DRED_Gamma5}
\{\gamma_5,\tilde\gamma_\mu\}=0;\qquad \gamma_5^2=1.
\end{equation}

\noindent
As a consequence it is possible obtain (see \cite{JackJones} for details),
that

\begin{equation}
(n-4)\,\mbox{tr}\Big(\gamma_5 \tilde\gamma_\mu
\tilde\gamma_\nu\tilde\gamma_\alpha\tilde\gamma_\beta\Big) = 0.
\end{equation}

Due to (\ref{DRED_Gamma5}) both gauge and chiral symmetries are
unbroken. In DREG this problem can be solved by use of $\gamma_5$ with the
following properties:

\begin{equation}
\{\gamma_5,\gamma_\mu\}=0,\quad \mu=0,\ldots,3;\qquad
[\gamma_5,\gamma_\mu]=0,\quad \mu>3,
\end{equation}

\noindent
which allow to derive the axial anomaly unambiguously \cite{tHV}.
Nevertheless (see review \cite{JackJones} and references therein)
DRED is usually believed to be a satisfactory regularization for
supersymmetric theories. From the practical point of view in order
to perform calculations by DRED it is necessary to use 4-dimensional
algebra of $\gamma$-matrices and calculate the remaining integrals
in the dimension $n\ne 4$ \cite{West}.


\section{Calculation of the two-loop $\beta$-function.}
\label{Section_Two_Loop}

\subsection{Two-loop Feynman diagrams.}
\hspace{\parindent}

\begin{figure}[p]
\hspace*{1.5cm}
\epsfxsize12.0truecm\epsfbox{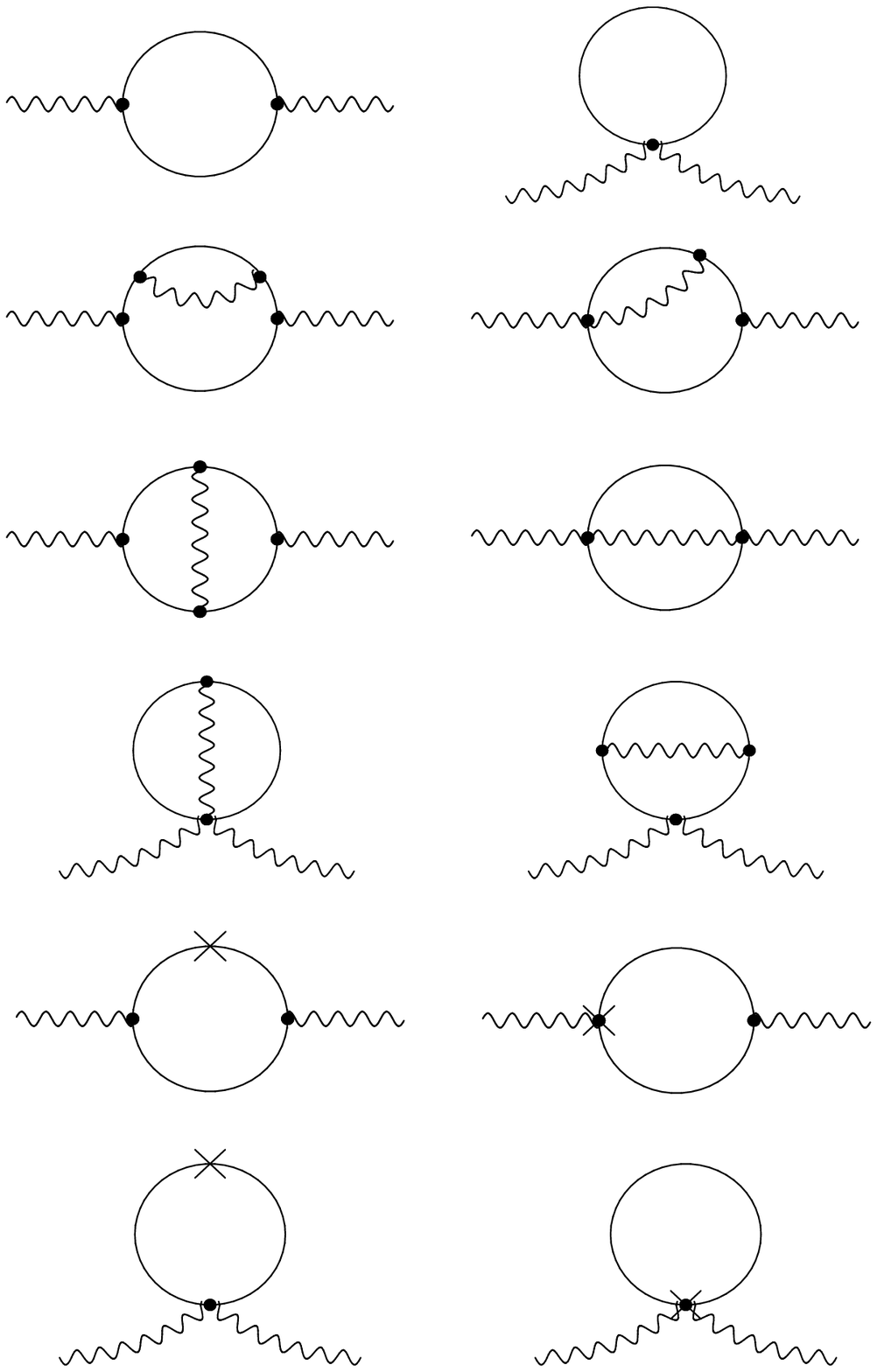}
\caption{Feynman diagrams giving nontrivial contributions to the
two-loop $\beta$-function.}
\begin{picture}(0,0)(0,0)
\put(1.5,20){$(1)$}
\put(8,20){$(2)$}
\put(1.5,17){$(3)$}
\put(8,17){$(4)$}
\put(1.5,14){$(5)$}
\put(8,14){$(6)$}
\put(1.5,10.5){$(7)$}
\put(8,10.5){$(8)$}
\put(1.5,7.5){$(9)$}
\put(8,7.5){$(10)$}
\put(1.5,4){$(11)$}
\put(8,4){$(12)$}
\end{picture}
\label{Figure_Beta_Diagrams}
\end{figure}

Feynman diagrams giving nontrivial contributions to the two-loop
$\beta$-function for $N=1$ supersymmetric electrodynamics are presented
at Figure \ref{Figure_Beta_Diagrams}. These diagrams can be naturally
divided into three parts:

1. one-loop diagrams (1) and (2);

2. two-loop diagrams (without subtraction diagrams) (3) -- (8);

3. subtraction diagrams (9) -- (12), containing insertions of one-loop
counterterms.


\subsection{Higher derivatives regularization.}
\hspace{\parindent}

Divergent part of the two-loop effective action for $N=1$ supersymmetric
electrodynamics was obtained in \cite{hep,tmf1}
\footnote{In \cite{tmf1} contributions of diagrams (1) -- (8) was
found correctly, but the contribution of diagrams (9) -- (12) was
omitted.}. The result is

\begin{equation}\label{Two_Loop_Effective_Action_HD}
\Delta\Gamma^{(2)}_{V} = \mbox{Re} \int d^2\theta\,
\frac{d^4p}{(2\pi)^4} W_a(p) C^{ab} W_b(-p)
\Big(f_1 + f_2 + f_{2PV} + f_3 \Big),
\end{equation}

\noindent
where

\begin{equation}\label{F1}
f_1 =
- \frac{i}{2}\Bigg(\int\frac{d^4k}{(2\pi)^4} \frac{1}{k^2 (k+p)^2}
- \sum\limits_i c_i \int\frac{d^4k}{(2\pi)^4}
\frac{1}{(k^2-M_i^2)\Big((k+p)^2-M_i^2\Big)}\Bigg)
\end{equation}

\noindent
is a one-loop result,

\begin{equation}\label{F2}
\ f_2 = - e^2 \int \frac{d^4k}{(2\pi)^4}\frac{d^4q}{(2\pi)^4}
\frac{(k+p+q)^2+q^2-k^2-p^2}{k^2\Big(1 + (-1)^n k^{2n}/\Lambda^{2n}\Big)
(k+q)^2 (k+p+q)^2 q^2 (q+p)^2}
\end{equation}

\noindent
is a sum of diagrams (3) -- (8) without contributions of Pauli-Villars
fields,

\begin{eqnarray}\label{F2PV}
&& f_{2PV} = e^2 \sum\limits_i c_i\,
\int \frac{d^4k}{(2\pi)^4}\frac{d^4q}{(2\pi)^4}\,
\frac{1}{k^2 \Big(1 + (-1)^n k^{2n}/\Lambda^{2n}\Big)}
\times\nonumber\\
&& \times
\Bigg[\frac{(k+p+q)^2+q^2-k^2-p^2}{
\Big((k+q)^2-M_i^2\Big) \Big((k+p+q)^2-M_i^2\Big) \Big(q^2-M_i^2\Big)
\Big((q+p)^2-M_i^2\Big)}
+\nonumber\\
&& \qquad\qquad\qquad\qquad\qquad\qquad
+ \frac{4 M_i^2}{\Big((k+q)^2-M_i^2\Big) \Big(q^2-M_i^2\Big)^2
\Big((q+p)^2-M_i^2\Big)} \Bigg] \qquad\quad
\end{eqnarray}

\noindent
is a contribution of diagrams (3) -- (8) with internal loop of
Pauli-Villars fields and

\begin{equation}\label{F3}
f_3 =
- \frac{i e^2}{2\pi^2}\ln\frac{\Lambda}{\mu}\,
\sum\limits_i c_i
\int \frac{d^4k}{(2\pi)^4}\,\frac{M_i^2}{(k^2-M_i^2)^2
\Big((k+p)^2-M_i^2\Big)}\qquad\quad
\end{equation}

\noindent
is a sum of diagrams (9) -- (12), which have insertions of the one-loop
counterterms.

According to the calculations performed in \cite{hep,tmf1} contribution
(\ref{F2PV}) is finite and does not affect divergent part of the effective
action. It is important, because some diagrams with Pauli-Villars loop
in principle can contain divergencies. However, Pauli-Villars
regularization always implies existence of divergent graphs, but these
divergences should be cancelled in the sum of all diagrams, that
actually takes place in the considered case.

The other contributions in (\ref{Two_Loop_Effective_Action_HD}) are
given by

\begin{eqnarray}
&& f_1 = \frac{1}{16\pi^2} \ln\frac{\Lambda}{p}+O(1);\nonumber\\
&& f_2 = \frac{1}{16\pi^2}\,\frac{\alpha}{\pi}\ln\frac{\Lambda}{p}+O(1);
\nonumber\\
&& f_3 = -\frac{1}{16\pi^2}\,\frac{\alpha}{\pi}\ln\frac{\Lambda}{\mu}
+o(1).
\end{eqnarray}

The corresponding two-loop contribution to the effective action is
($\Gamma_R = S_{ren}+\Delta\Gamma$)

\begin{equation}\label{2-Loop_Effective_Action}
\Delta\Gamma^{(2)}_{V} = \frac{1}{16\pi^2}\mbox{Re} \int d^2\theta\,
\frac{d^4p}{(2\pi)^4} W_a(p) C^{ab} W_b(-p)
\Bigg(\ln\frac{\Lambda}{p}
+\frac{\alpha}{\pi}\ln\frac{\mu}{p}+O(1)\Bigg),
\end{equation}

\noindent
so that it is not necessary to add any new counterterms for cancellation
of the two-loop divergencies:

\begin{eqnarray}
&& \Delta S =  - \frac{1}{16\pi^2}\mbox{Re} \int d^2\theta\,
\frac{d^4p}{(2\pi)^4} W_a(p) C^{ab} W_b(-p)
\Bigg(
\ln\frac{\Lambda}{\mu}
+ \mbox{finite terms} + O(\alpha^2)
\Bigg)
+\quad\nonumber\\
&& + \mbox{terms with matter superfields},
\end{eqnarray}

\noindent
so that

\begin{equation}
\frac{1}{\alpha_0} = \frac{1}{\alpha\Big(\Lambda/\mu\Big)}
- \frac{1}{\pi} \ln \frac{\Lambda}{\mu} + O(\alpha^2).
\end{equation}

\noindent
According to equation (\ref{Beta_Gamma_Definition}) this expression
corresponds to the following two-loop $\beta$-function:

\begin{equation}\label{Beta_HD}
\beta_{HD} = \frac{\alpha^2}{\pi} + O(\alpha^4).
\end{equation}


\subsection{Dimensional reduction.}
\hspace{\parindent}

A result corresponding to (\ref{Two_Loop_Effective_Action_HD}),
obtained by DRED, is given by

\begin{equation}\label{Two_Loop_Effective_Action_DRED}
\Delta\Gamma^{(2)}_{V} = \mbox{Re} \int d^2\theta\,
\frac{d^4p}{(2\pi)^4} W_a(p) C^{ab} W_b(-p)
\Big(\tilde f_1 + \tilde f_2 + \tilde f_3 \Big),
\end{equation}

\noindent
where

\begin{eqnarray}\label{TildeF1}
&& \tilde f_1 =
- \frac{i}{2}\int\frac{d^nk}{(2\pi)^4} \frac{1}{k^2 (k+p)^2};\\
\label{TildeF2}
&& \tilde f_2 = - e^2 \int \frac{d^nk}{(2\pi)^4}\frac{d^nq}{(2\pi)^4}
\frac{(k+p+q)^2+q^2-k^2-p^2}{k^2 (k+q)^2 (k+p+q)^2 q^2 (q+p)^2};\\
\label{TildeF3}
&& \tilde f_3 = 0.\vphantom{\frac{1}{2}}
\end{eqnarray}

\noindent
Using notations of \cite{Leveille}

\begin{eqnarray}
&& I \equiv (p^2)^{2-n/2} \int \frac{d^nk}{k^2(k+p)^2}
=\pi^{n/2}\Gamma\Big(2-n/2\Big) B\Big(n/2-1,n/2-1\Big);
\nonumber\\
&& J \equiv (p^2)^{4-n} \int \frac{d^nk}{(k^2)^{3-n/2}(k+p)^2}
= \pi^{n/2}\frac{\Gamma\Big(4-n\Big)}{\Gamma\Big(3-n/2\Big)}
B\Big(n-3,n/2-1\Big);\nonumber\\
&& Z \equiv (p^2)^{5-n}
\int \frac{d^nk\,d^nq}{q^2 k^2 (k+p)^2 (q+p)^2 (k-q)^2}
= \frac{1}{n-4}\Big((6n-20) IJ -(2n-6) I^2\Big)\nonumber\\
\end{eqnarray}

\noindent
the two-loop contribution to the effective action, calculated by DRED,
can be written as

\begin{eqnarray}
&& \Delta\Gamma^{(2)}_{V} = \mbox{Re} \int d^2\theta\,
\frac{d^4p}{(2\pi)^4} W_a(p) C^{ab} W_b(-p)
\times\nonumber\\
&& \qquad\qquad\qquad
\times
\Bigg(\frac{1}{2 (2\pi)^4} (p/\mu_0)^{n-4} I
+ \frac{e^2}{(2\pi)^8}(p/\mu_0)^{2n-8}\Big(2 IJ - I^2 - p^2 Z\Big) \Bigg),
\qquad
\end{eqnarray}

\noindent
where a constant $\mu_0$ is present because the dimension of the
coupling constant depends on the space-time dimension $n$ \cite{MS}.

Having calculated the integrals we obtain, that

\begin{eqnarray}\label{Gamma_DRED}
&& \Delta\Gamma^{(2)}_{V} = \frac{1}{16\pi^2}\mbox{Re} \int d^2\theta\,
\frac{d^4p}{(2\pi)^4} W_a(p) C^{ab} W_b(-p)
\times\nonumber\\
&& \qquad\qquad\qquad\quad
\times \Bigg(
\frac{1}{4-n}+\ln\frac{\mu_0}{p}
+ \frac{\alpha}{\pi} \Bigg(\frac{1}{2(4-n)}+\ln \frac{\mu_0}{p}\Bigg)
+ O(1) \Bigg).
\qquad
\end{eqnarray}

\noindent
First two terms here correspond to the one-loop integral $\tilde f_1$,
and the third term corresponds to the two-loop integral $\tilde f_2$.

In order to cancel divergences in $\Delta\Gamma^{(2)}_V$ it is necessary
to add counterterms

\begin{eqnarray}
&& \Delta S =  \frac{1}{16\pi^2}\mbox{Re} \int d^2\theta\,
\frac{d^4p}{(2\pi)^4} W_a(p) C^{ab} W_b(-p)
\times\nonumber\\
&& \qquad
\times \Bigg(
-\frac{1}{4-n}+\ln\frac{\mu}{\mu_0}
+ \frac{\alpha}{\pi} \Bigg(-\frac{1}{2(4-n)}+\ln \frac{\mu}{\mu_0}\Bigg)
+ \mbox{finite terms} + O(\alpha^2)
\Bigg).\qquad
\end{eqnarray}

\noindent
Therefore the DRED $\beta$-function defined by
(\ref{Beta_Gamma_Definition}) is equal to

\begin{equation}\label{Beta_DRED}
\beta_{DRED} = \frac{\alpha^2}{\pi} + \frac{\alpha^3}{\pi^2} + O(\alpha^4),
\end{equation}

\noindent
while the renormalized effective action coincides with the one, obtained
by HD regularization. Unlike (\ref{Beta_HD}) expression (\ref{Beta_DRED})
agrees with the NSVZ $\beta$-function (\ref{NSVZ_Beta}).


\subsection{Comparison between HD and DRED regularizations.}
\hspace{\parindent}

Comparing the calculations, described above, we see that the difference
of the $\beta$-functions comes from the different results for the sum of
diagrams with insertions of one-loop counterterms. In order to understand
why these results are different it is necessary to note, that the sum of
diagrams with insertions of counterterms in the considered approximation
is equal to Konishi anomaly \cite{Konishi,ClarkKonishi}. The existence
of Konishi anomaly can be explained by following arguments:

Let us consider the following expression:

\begin{equation}\label{Im_Of_Konishi}
\mbox{Im}\Bigg[\bar D^2 \Big(\phi^* e^{2V}\phi
+ \tilde\phi^* e^{-2V}\tilde\phi\Big)\Bigg].
\end{equation}

\noindent
Using equation (\ref{Phi_Superfield}) and (\ref{V_Superfield}), it is
easy to see, that in components it will contain (among other terms)

\begin{equation}\label{Axial_Current}
- \bar\theta\theta\,
\partial_\mu \Big(\bar\Psi \gamma^\mu\gamma_5\Psi\Big),
\end{equation}

\noindent
where the Dirac spinor $\Psi$ is defined by (\ref{Psi_Definition}).
It is well known \cite{Bertlmann}, that the conservation of the axial
current is broken by quantum corrections and in particular

\begin{equation}\label{Axial_Anomaly}
\langle\,\bar\theta\theta\,
\partial_\mu \Big(\bar\Psi \gamma^\mu\gamma_5\Psi\Big)\rangle
= - \bar\theta\theta\,\frac{1}{8\pi^2} F_{\mu\nu}\tilde F^{\mu\nu}.
\end{equation}

\noindent
Hence due to the supersymmetry
\footnote{Note, that our arguments can not be considered as a derivation
of Konishi anomaly, because (\ref{Axial_Current}) does not contain
all terms of (\ref{Im_Of_Konishi}), proportional to $\bar\theta\theta$.
A strict derivation of the Konishi anomaly can be found in
\cite{Konishi,ClarkKonishi}. Our goal is only to remind relation between
axial anomaly and Konishi anomaly.}

\begin{equation}\label{Im_Of_Anomaly}
\mbox{Im}\,\Big\langle \bar D^2 \Big(\phi^* e^{2V}\phi
+ \tilde\phi^* e^{-2V}\tilde\phi\Big)\Big\rangle
= \frac{1}{2\pi^2} \mbox{Im}\Big(W_a C^{ab} W_b\Big).
\end{equation}

\noindent
Performing supersymmetry transformations it is easy to see, that if
imaginary part of a chiral superfield is equal to 0, then this superfield
is a real constant. Therefore, from (\ref{Im_Of_Anomaly}) we obtain, that

\begin{equation}\label{Full_Anomaly}
\Big\langle \bar D^2 \Big(\phi^* e^{2V}\phi
+ \tilde\phi^* e^{-2V}\tilde\phi\Big)\Big\rangle
= \frac{1}{2\pi^2} W_a C^{ab} W_b + \mbox{const}.
\end{equation}

\noindent
Applying

\begin{equation}
-\frac{1}{2}\int d^4x\, D^2 = \int d^4x\,d^2\theta
\end{equation}

\noindent
to (\ref{Full_Anomaly}) and taking a real part of the result, we obtain,
that

\begin{equation}\label{Konishi_Anomaly}
\Big\langle \frac{1}{4}\int d^4x\,d^4\theta\,\Big(\phi^* e^{2V}\phi
+ \tilde\phi^* e^{-2V}\tilde\phi\Big)\Big\rangle
= -\frac{1}{16\pi^2} \mbox{Re}\int d^4x\,d^2\theta\,W_a C^{ab} W_b.
\end{equation}

It is important to note, that if the axial anomaly is found to be equal
to 0 due to some reason and the supersymmetry is unbroken, instead of
(\ref{Konishi_Anomaly}) we will automatically obtain

\begin{equation}
\Big\langle \frac{1}{4}\int d^4x\,d^4\theta\,\Big(\phi^* e^{2V}\phi
+ \tilde\phi^* e^{-2V}\tilde\phi\Big)\Big\rangle = 0.
\end{equation}

\noindent
This takes place if the calculations are made by DRED. Really, if
we calculate supergraphs, then it is impossible to impose requirements
similar to $\mbox{tr}(AB)\ne\mbox{tr}(BA)$. Therefore, it is necessary
to use original variant of DRED, proposed in \cite{Siegel} with $n<4$
and $\gamma_5$, satisfying (\ref{DRED_Gamma5}). It means, that the chiral
symmetry is not broken in the regularized theory and anomaly is equal to
0 due to the mathematical inconsistency of DRED. As a consequence Konishi
anomaly is also equal to 0, that, in turn, leads to zero result for the
sum of diagrams with insertions of counterterms. Taking into account that
in the two-loop approximation these diagrams should cancel the other
contributions, DRED leads to a nontrivial two-loop correction and to the
anomaly puzzle.

%
%
%
%


\section{Exact result for diagrams with insertions of counterterms.}
\hspace{\parindent}
\label{Section_Exact}

In the previous section we found, that the difference between HD and
DRED $\beta$-functions originated from different results for the sum
of diagrams with insertions of counterterms. In DRED this sum is equal
to zero, because this regularization does not allow to calculate
Konishi anomaly, which can be found correctly by HD regularization.
In order to confirm such arguments in this section a sum of diagrams with
insertion of counterterms is calculated exactly to all orders using
HD regularization.

The result obtained in appendix can be written in the following form:

\begin{eqnarray}\label{Result_For_Diagrams_With_Counterterms}
&& \exp\Big(i\Gamma\Big) = \exp\Bigg\{-i\ln Z\,\frac{1}{16\pi^2}\,
\mbox{Re}\int d^4x\,d^2\theta\,W_a C^{ab} W_b
+\mbox{finite terms}\Bigg\}\,
\times\nonumber\\
&& \times
\int DV\,D\phi\,D\tilde \phi\,
\prod\limits_i \Big(\det{}' PV(V,M_i)\Big)^{c_i}\,
\exp\Bigg\{i\Bigg[\frac{1}{4 e^2} \int d^4x\,d^4\theta\, V\partial^2
\Big(1+ \frac{\partial^{2n}}{\Lambda^{2n}}\Big) V
-\nonumber\\
&& - \frac{1}{4 e^2} \Big(Z_3(\Lambda/\mu)-1\Big) \int d^4x\,d^4\theta\,
V \Pi_{1/2}\partial^2
\Big(1+ \frac{\partial^{2n}}{\Lambda^{2n}}\Big) V
+\nonumber\\
&& \qquad\qquad\qquad\qquad\qquad\qquad\qquad\qquad
+ \frac{1}{4} \int d^4x\,d^4\theta\,
\Big(\phi^* e^{2V}\phi
+ \tilde\phi^* e^{-2V}\tilde\phi \Big)
+\nonumber\\
&&
+ \int d^4x\,d^4\theta\,J V
+ \int d^4x\,d^2\theta\, \Big(j\,\phi + \tilde j\,\tilde\phi \Big)
+ \int d^4x\,d^2\bar\theta\,
\Big(j^*\phi^* + \tilde j^* \tilde\phi^* \Big)\Bigg]\Bigg\},
\end{eqnarray}

\noindent
where $J$, $j$ and $\tilde j$ should be eliminated in terms of $V$,
$\phi$ and $\tilde\phi$ and

\begin{eqnarray}
&& \Big(\det{}' PV(V,M)\Big)^{-1} \equiv \int D\Phi\,D\tilde \Phi\,
\exp\Bigg\{i\Bigg[\frac{1}{4} \int d^4x\,d^4\theta\,
\Big(\Phi^* e^{2V}\Phi
+\qquad\nonumber\\
&& + \tilde\Phi^* e^{-2V}\tilde\Phi \Big)
+ \frac{1}{2}\int d^4x\,d^2\theta\, M \tilde\Phi \Phi
+ \frac{1}{2}\int d^4x\,d^2\bar\theta\, M \tilde\Phi^* \Phi^*
\Bigg]\Bigg\}.
\end{eqnarray}

\noindent
(The difference between this definition and (\ref{PV_Determinants})
is the absence of $Z$ in definition of $\det{}'$.)

Equation (\ref{Result_For_Diagrams_With_Counterterms}) can be formally
written in the more simple form:

\begin{eqnarray}\label{Generalization_Of_Konishi_Anomaly}
&& \Big\langle \exp\Bigg(i (Z-1)\,
\frac{1}{4} \int d^4x\,d^4\theta\,\Big(\phi^* e^{2V}\phi
+ \tilde\phi^* e^{-2V}\tilde\phi\Big)\Bigg)\Big\rangle
=\nonumber\\
&& \qquad\qquad\qquad\quad
= \exp\Bigg(-i\ln Z
\frac{1}{16\pi^2} \mbox{Re}\int d^4x\,d^2\theta\,W_a C^{ab} W_b
+\mbox{finite terms}\Bigg),
\qquad
\end{eqnarray}

\noindent
Therefore, althought expression
(\ref{Result_For_Diagrams_With_Counterterms}) is rather complicated,
its essence is quite simple: rescaling of $\phi$ and $\tilde\phi$
produces a factor

\begin{equation}
\exp\Bigg\{-i\ln Z\,\frac{1}{16\pi^2}\,
\mbox{Re}\int d^4x\,d^2\theta\,W_a C^{ab} W_b
+\mbox{finite terms}\Bigg\}.
\end{equation}

\noindent
From (\ref{Result_For_Diagrams_With_Counterterms}) we conclude, that
the sum of diagrams with insertions of counterterms gives the following
contribution to the $\beta$-function:

\begin{equation}
\Delta\beta = \frac{\alpha^2}{\pi}\,\frac{d\ln Z}{d\ln\mu}
= \frac{\alpha^2}{\pi}\gamma(\alpha).
\end{equation}

\noindent
Therefore, if the $\beta$-function obtained by HD regularizarion
is defined by the one-loop approximation, the sum of all diagrams
without insertions of counterterms on the matter lines will give
NSVZ $\beta$-function (\ref{NSVZ_Beta}). Such result can be obtain
after rescaling of matter superfields $\phi\to Z^{-1/2}\phi$, which
convert renormalized action (\ref{Renormalized_Action}) to

\begin{eqnarray}\label{Renormalized_Action_After_Rescaling}
&& S_{ren} =
\frac{1}{4 e^2} Z_3(\Lambda/\mu) \mbox{Re}\int d^4x\,d^2\theta\,W_a C^{ab}
\Big(1+ \frac{\partial^{2n}}{\Lambda^{2n}}\Big) W_b
+\nonumber\\
&& \qquad\qquad\qquad\qquad\qquad\qquad
+ \frac{1}{4}\int d^4x\, d^4\theta\,
\Big(\phi^* e^{2V}\phi +\tilde\phi^* e^{-2V}\tilde\phi\Big).\qquad
\end{eqnarray}

\noindent
Then the diagrams with insertions of counterterms will be evidently absent.
Such possibility was investigated in \cite{Arkani} by other methods.

Another possibility to obtain (\ref{NSVZ_Beta}) is using of DRED.
In this case the sum of diagrams with insertions of counterterms
is calculated incorrectly and is equal to 0. As a consequence we
obtain anomaly puzzle, which appears as an artifact of mathematical
inconsistency of DRED. However, the renormalized effective action is
the same for both regularizations and the only difference between
these regularizations is a value of the $\beta$-function.

Note, that if HD regularization leads to the one-loop $\beta$-function
\cite{hep}, the result obtained in DRED will give the NSVZ $\beta$-function
after redefinition of the coupling constant, that is in agreement with
the results of \cite{Jones,North}.

Finally it is necessary to mention, that equations similar to
(\ref{Generalization_Of_Konishi_Anomaly}) were also obtained
in \cite{SV} and \cite{Arkani}. However anomalous contribution in
(\ref{Result_For_Diagrams_With_Counterterms}) was not identified with
the sum of diagrams with insertions of counterterms. Nevertheless,
the derivation of this equation presented in \cite{SV} is in a certain
degree similar to the derivation, given in this paper.


\section{Conclusion.}
\hspace{\parindent}
\label{Section_Conclusion}

In this paper calculation of the two-loop $\beta$-function for $N=1$
supersymmetric electrodynamics was analyzed for regularizations by DRED
and HD. Now let us summarize the results.

1. Konishi anomaly really gives a nontrivial contribution to the
$\beta$-function, as it was pointed out in \cite{SV}. This anomaly
can be identified with the sum of Feynman diagrams with insertions
of counterterms. Nevertheless, contribution of these diagrams does
not change form of the renormalized effective action.

2. Existence of rescaling anomaly, investigated in \cite{Arkani}, can be
easily explained by the diagram technique: Diagrams with insertions
of counterterms (on lines of matter superfields) are present only if
the renormalized action contains

\begin{equation}
\frac{1}{4} Z \int d^4x\,d^4\theta\,\Big(\phi^* e^{2V}\phi
+ \tilde\phi^* e^{-2V} \tilde\phi\Big),
\end{equation}

\noindent
that corresponds to the holomorphic normalization of matter superfields.
After rescaling $\phi\to Z^{-1/2}\phi$ such diagrams disappear. Therefore,
due to the Konishi anomaly these transformations will be anomalous.

3. If the $\beta$-function is calculated by a mathematically consistent
regularization, which does not break supersymmetry, then the anomaly puzzle
is absent (at least at the two-loop level). This means, that the
$\beta$-function is completely defined by the one-loop approximation.
An example of such regularization is regularization by HD, considered in
this paper. It is important, that HD regularization allows to obtain
an exact expression for the sum of subtraction diagrams, which agrees
with the results found in \cite{SV} and \cite{Arkani} from other arguments.
Possibly the NSVZ $\beta$-function can be obtained if one uses HD
regularization and subtractions at some scale $\mu$, as it was proposed
in \cite{hep}. However, this can be checked only at the three loops,
because starting from the three-loop approximation coefficients of the
$\beta$-function depend on the renormalization scheme. This work has been
already finished and the corresponding paper is in preparation.

4. Ambiguities of DRED lead to the incorrect result for the axial anomaly
and, as a consequence, for the Konishi anomaly. This, in turn, produces
incorrect $\beta$-function and the anomaly puzzle. At the present moment
I do not know how to impose requirements like
$\mbox{tr}(AB)\ne \mbox{tr}(BA)$, allowing to find anomalies by DRED,
within the explicitly supersymmetric technique of calculations.


It is necessary to note, that so far we considered only the Abelian case.
For the supersymmetric Yang-Mills theory using of higher covariant
derivative regularization \cite{West_Paper} leads to very involved
calculations, because in this case Feynman rules become much more
complicated. In this case using of usual derivatives can considerably
simplify the calculations. However, such regularization breaks the gauge
invariance. Nevertheless, even in case of noninvariant regularization it
is possible to obtain gauge invariant renormalized effective action by
a special choice of subtraction scheme \cite{Slavnov1,Slavnov2}.
For Abelian supersymmetric theories such scheme was proposed in
\cite{SlavnovStepan1}. Construction of invariant renormalization
procedure for supersymmetric non-Abelian models is in progress.

Also I would like to mention, that NSVZ $\beta$-function was obtained
not only in DRED, but also in the differential renormalization (DiffR)
\cite{DiffR}. The calculations were made in \cite{Mas} for $N=1$
supersymmetric Yang-Mills theory. At present I can not trace the
origin of the multiloop corrections in this case and believe, that it
would be interesting to compare the calculation of this paper with the
corresponding results, obtained in DiffR.


\vspace{1cm}

\noindent
{\Large{\bf Acknowledgments}}

\bigskip

I would like to thank V.A.Novikov, P.I.Pronin, A.A.Slavnov and
A.A.Soloshenko for valuable discussions.


\vspace{1cm}

\noindent
{\Large\bf Appendix.}


\appendix


\section{Exact sum of diagrams with insersions of counterterms.}
\hspace{\parindent}
\label{Appendix_Counterterms}

In order to prove (\ref{Result_For_Diagrams_With_Counterterms}) it is
necessary to consider 1PI diagrams with insertions of counterterms on
lines of the matter superfield. A sum of diagrams, containing internal
$V$-lines, is finite due to the HD regularization. Hence, nontrivial
contributions can be given only by diagrams with a single loop of matter
superfields without internal $V$-lines. A sum of such contributions can
be calculated exactly. Really, it is easy to see, that the effective line

\bigskip

\begin{equation}\label{Effective_Line_Picture}
\qquad\qquad\qquad\quad
\smash{\epsfxsize6.0truecm\epsfbox[240 392 440 892]{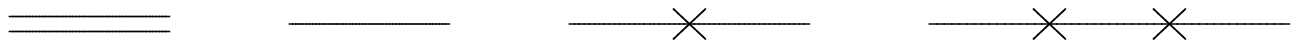}}
\hspace*{-8.6cm} =
\hspace*{1.9cm} +
\hspace*{2.8cm} +
\hspace*{3.7cm} + \ \ \ldots,
\end{equation}

\bigskip

\noindent
where crosses denote contributions from counterterms, corresponds
to the following propagators:

\begin{equation}
\frac{1}{Z}\left(
\begin{array}{cccc}
0 & {\displaystyle -\frac{\bar D^2 D^2}{16 (\partial^2 + m^2/Z^2)}} &
{\displaystyle \frac{m\bar D^2}{4 Z(\partial^2+m^2/Z^2)}
\vphantom{\int\limits_p}} & 0 \\
{\displaystyle - \frac{D^2\bar D^2}{16(\partial^2 + m^2/Z^2)}} & 0 & 0 &
{\displaystyle \frac{m D^2}{4Z(\partial^2+m^2/Z^2)}
\vphantom{\int\limits_p}} \\
{\displaystyle \frac{m\bar D^2}{4Z(\partial^2+m^2/Z^2)}
\vphantom{\int\limits_p}} & 0 & 0 &
{\displaystyle \frac{\bar D^2 D^2}{16 (\partial^2 + m^2/Z^2)}} \\
0 & {\displaystyle  \frac{m D^2}{4Z(\partial^2+m^2/Z^2)}
\vphantom{\int\limits_p}}
& {\displaystyle \frac{D^2\bar D^2}{16(\partial^2+m^2/Z^2)}} & 0
\end{array}
\right)
\end{equation}

\noindent
($m=0$ for $\phi$ and $\tilde\phi$ lines and $m=M$ for lines of
Pauli-Villars fields.) The first string of this matrix corresponds
to the propagators $\phi-\phi$, $\phi-\phi^*$, $\phi-\tilde\phi$,
$\phi-\tilde\phi^*$, the second string corresponds to the propagators
$\phi^*-\phi$, $\phi^*-\phi^*$, $\phi^*-\tilde\phi$, $\phi^*-\tilde\phi^*$
e t.c.

These expressions should be substituted to the diagrams

\bigskip
\bigskip

\begin{equation}\label{One-Loop_Effective_Diagrams_Picture}
\qquad \smash{\epsfxsize6.0truecm\epsfbox[190 390 490 890]{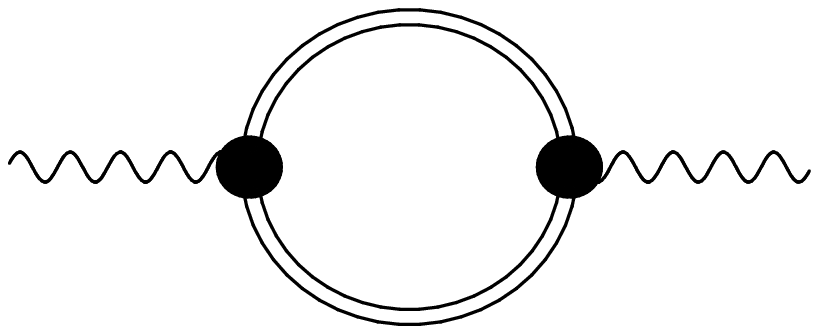}}
\hspace{-1cm}
+ \hspace{-0.5cm}
\smash{\epsfxsize6.0truecm\epsfbox[190 370 490 870]{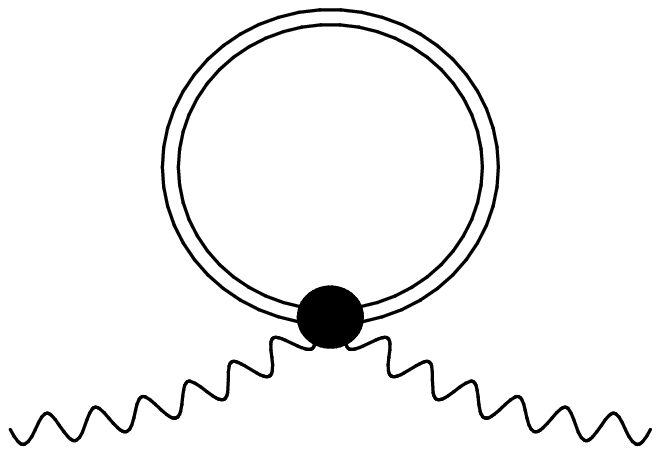}}
\end{equation}

\bigskip
\bigskip

\noindent
where circles denotes the effective vertexes

\bigskip
\bigskip
\bigskip

\begin{equation}
\qquad \smash{\epsfxsize6.0truecm\epsfbox[140 390 440 890]{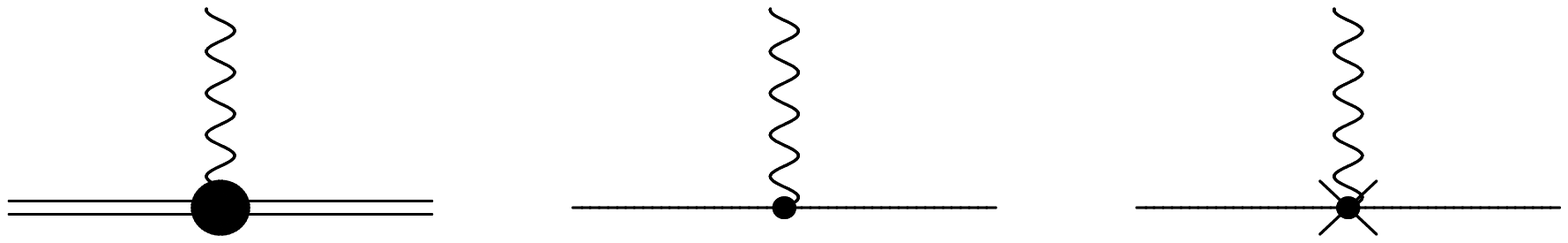}}
\hspace*{-4.8cm} =
\hspace*{3.25cm} +
\qquad\qquad\qquad\quad
\end{equation}

\bigskip

\noindent
and

\bigskip

\begin{equation}
\qquad \smash{\epsfxsize6.0truecm\epsfbox[140 390 440 890]{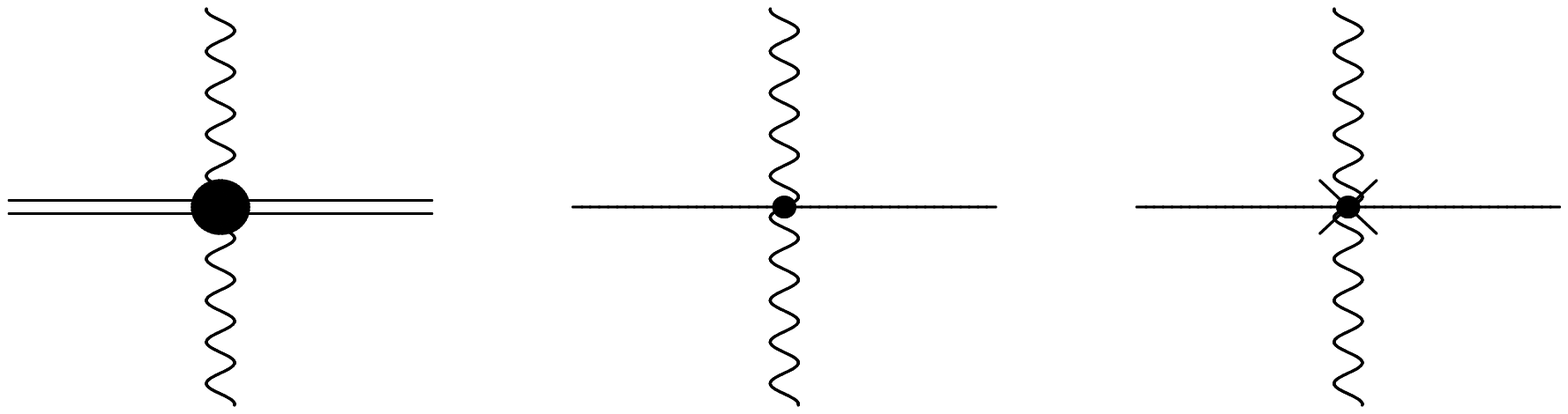}}
\hspace*{-4.8cm} =
\hspace*{3.25cm} +
\qquad\qquad\qquad\quad
\end{equation}

\bigskip
\bigskip

\noindent
which are proportional to

\begin{equation}
Z\left(
\begin{array}{cccc}
0 & 1 & 0 & 0 \\
1 & 0 & 0 & 0 \\
0 & 0 & 0 & -1 \\
0 & 0 & -1 & 0
\end{array}
\right)
\quad\mbox{and}\quad
Z\left(
\begin{array}{cccc}
0 & 1 & 0 & 0 \\
1 & 0 & 0 & 0 \\
0 & 0 & 0 & 1 \\
0 & 0 & 1 & 0
\end{array}
\right)
\end{equation}

\noindent
respectively. Then it is quite evident, that the expression for
diagrams (\ref{One-Loop_Effective_Diagrams_Picture}) will differ
from the corresponding expression for usual one-loop diagrams

\bigskip
\bigskip
\bigskip

\begin{equation}\label{One-Loop_Diagrams_Picture}
\qquad \smash{\epsfxsize6.0truecm\epsfbox[190 390 490 890]{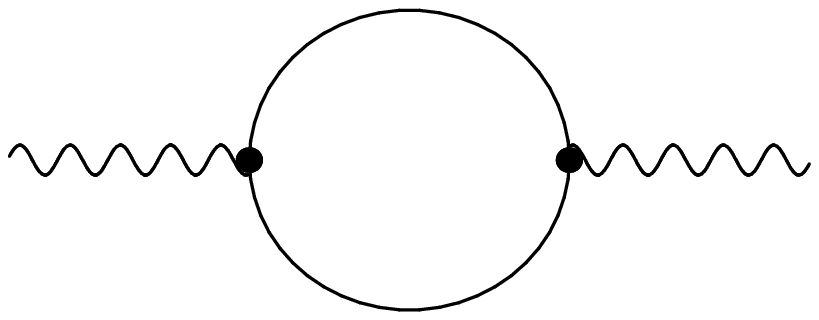}}
\hspace{-1cm}
+ \hspace{-0.5cm}
\smash{\epsfxsize6.0truecm\epsfbox[190 370 490 870]{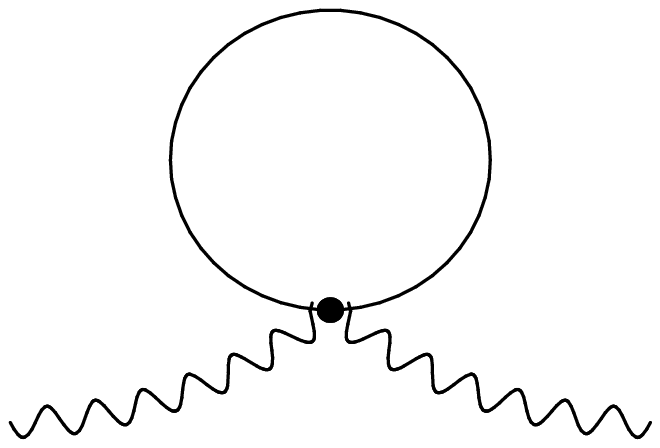}}
\end{equation}

\bigskip
\bigskip
\bigskip

\noindent
by the substitution $M\to M/Z$ and is equal to

\begin{eqnarray}
&& - \frac{i}{2}\mbox{Re} \int d^2\theta \frac{d^4p}{(2\pi)^2}\,
W_a(p,\theta) C^{ab} W_b(-p,\theta)\,
\times\nonumber\\
&& \qquad\quad
\times
\int \frac{d^4k}{(2\pi)^2}\Bigg(\frac{1}{k^2 (k+p)^2}
-\sum\limits_i c_i
\frac{1}{\Big(k^2-M_i^2/Z^2\Big)\Big((k+p)^2-M_i^2/Z^2\Big)}\Bigg).
\qquad
\end{eqnarray}

\noindent
Subtracting the result for the one-loop diagrams from this expression
we obtain the following result for the diagrams with insertions of
counterterms:

\begin{eqnarray}
&& - \frac{i}{2}\mbox{Re} \int d^2\theta \frac{d^4p}{(2\pi)^2}\,
W_a(p,\theta) C^{ab} W_b(-p,\theta)\,
\times\vphantom{\Bigg(}\nonumber\\
&& \qquad\qquad\quad
\times
\sum\limits_i c_i
\int \frac{d^4k}{(2\pi)^2}\Bigg(
\smash{\frac{1}{\Big(k^2-M_i^2\Big)\Big((k+p)^2-M_i^2\Big)}}
-\nonumber\\
&& \qquad\qquad\qquad\qquad\qquad\qquad\qquad\quad
- \frac{1}{\Big(k^2-M_i^2/Z^2\Big)\Big((k+p)^2-M_i^2/Z^2\Big)}
\Bigg).\qquad
\end{eqnarray}

Performing Wick rotation and taking into account that

\begin{eqnarray}
&& \frac{1}{2}\int \frac{d^4k}{(2\pi)^4}
\frac{1}{\Big(k^2+M^2/Z^2\Big)\Big((k+p)^2+M^2/Z^2\Big)}
-\nonumber\\
&& \qquad\qquad\qquad\qquad\qquad\qquad\qquad
- \frac{1}{2}\int \frac{d^4k}{(2\pi)^4}
\frac{1}{\Big(k^2 +M^2\Big)\Big((k+p)^2+M^2\Big)}
=\nonumber\\
&& = -\frac{1}{16\pi^2} \Bigg(
\ln\frac{M}{p Z}
+ \sqrt{1+\frac{4M^2}{p^2 Z^2}}\,
\mbox{arctanh}\sqrt{\frac{p^2}{4M^2/Z^2 + p^2}}
-\nonumber\\
&& \qquad\qquad\qquad\qquad\qquad\qquad\quad
- \ln\frac{M}{p}
- \sqrt{1+\frac{4M^2}{p^2}}\,
\mbox{arctanh}\sqrt{\frac{p^2}{4M^2 + p^2}}\Bigg)
=\qquad\nonumber\\
&& = \frac{1}{16\pi^2} \ln Z+\mbox{finite terms},\qquad\ \
\end{eqnarray}

\noindent
the sum of the considered 1PI diagrams with insertions of counterterms
appears to be

\begin{equation}\label{1PI_With_Counterterms}
- \ln Z\,\frac{1}{16\pi^2}\,
\mbox{Re}\int d^4x\,d^2\theta\,W_a C^{ab} W_b +\mbox{finite terms}.
\end{equation}

\noindent
It is easy to see, that in terms of the generating functional this
equation can be written as (\ref{Result_For_Diagrams_With_Counterterms}).


\pagebreak

\begin{thebibliography}{100}

\bibitem{Ferrara}
{\it S.Ferrara, B.Zumino}, Nucl.Phys. {\bf B 87}, (1975), 207.

\bibitem{Clark}
{\it T.E.Clark, O.Piquet, K.Sibold}, Nucl.Phys. {\bf B 143}, 445, (1978).

\bibitem{Piquet1}
{\it O.Piquet, K.Sibold}, Nucl.Phys. {\bf B 196}, 428, (1982).

\bibitem{Piquet2}
{\it O.Piquet, K.Sibold}, {\it Nucl.Phys.} {\bf B 196}, 447, (1982).

\bibitem{Bardeen}
{\it S.L.Adler, W.A.Bardeen}, Phys.Rev. {\bf 182}, 1517, (1969).

\bibitem{Slavnov_Book}
{\it L.D.Faddeev, A.A.Slavnov}, Gauge fields, introduction to quantum
theory, second edition, Benjamin, Reading, 1990.

\bibitem{Adler_Collins}
{\it S.L.Adler, J.C.Collins, A.Duncan}, Phys.Rev. {\bf D 15}, 1712, (1977).

\bibitem{NSVZ_PL}
{\it V.A.Novikov, M.A.Shifman, A.I.Vainstein, V.I.Zakharov}, Phys.Lett.
{\bf 157B}, (1985), 169.

\bibitem{N2}
{\it P.S.Howe, K.S.Stelle, P.S.West}, Phys.Lett. {\bf 124B}, (1983), 55.

\bibitem{Tarasov}
{\it O.V.Tarasov, V.A.Vladimirov}, Phys.Lett. {\bf 96 B}, 94, (1980).

\bibitem{Grisaru}
{\it M.T.Grisaru, M.Rocek, W.Siegel}, Phys.Rev.Lett. {\bf 45}, 1063,
(1980).

\bibitem{Caswell}
{\it W.Caswell, D.Zanon}, Phys.Lett. {\bf 100 B}, 152, (1980).

\bibitem{SV}
{\it M.A.Shifman, A.I.Vainstein}, Nucl.Phys. {\bf B277}, (1986), 456.

\bibitem{Konishi}
{\it K.Konishi}, Phys.Lett. {\bf 135B}, (1984), 439.

\bibitem{ClarkKonishi}
{\it T.E.Clark, O.Piquet, K.Sibold}, Nucl.Phys. {\bf B159}, (1979), 1.

\bibitem{NSVZ_Instanton}
{\it V.A.Novikov, M.A.Shifman, A.I.Vainstein, V.I.Zakharov}, Phys.Lett.
{\bf 166B}, (1985), 329.

\bibitem{ThreeLoop1}
{\it L.V.Avdeev, O.V.Tarasov}, Phys.Lett. {\bf 112 B}, (1982), 356.

\bibitem{ThreeLoop2}
{\it I.Jack, D.R.T.Jones, C.G.North}, Nucl.Phys. {\bf B 473}, (1996), 308

\bibitem{ThreeLoop3}
{\it I.Jack, D.R.T.Jones, C.G.North}, Phys.Lett. {\bf 386 B}, (1996), 138

\bibitem{Jones}
{\it I.Jack, D.R.T.Jones, C.G.North}, N=1 supersymmetry and the three loop
anomalous dimension for the chiral superfield, hep-ph/9603386.

\bibitem{JackJones}
{\it I.Jack, D.R.T.Jones}, Regularization of supersymmetric theories,
hep-ph/ 9707278.

\bibitem{North}
{\it I.Jack, D.R.T.Jones, C.G.North}, Nucl.Phys. {\bf B 486}, (1997), 479.

\bibitem{Arkani}
{\it N.Arkani-Hamed, H.Mirayama}, JHEP {\bf 0006}, (2000), 030,
(hep-th/9707133).

\bibitem{Slavnov}
{\it A.A.Slavnov}, Theor.Math.Phys. {\bf 23}, (1975), 3.

\bibitem{Bakeyev}
{\it T.Bakeyev, A.A.Slavnov}, Mod.Phys.Lett. {\bf A11}, (1996), 1539.

\bibitem{PhysLett}
{\it P.I.Pronin, K.V.Stepanyantz}, Phys.Lett. {\bf B414}, (1997), 117.

\bibitem{hep}
{\it A.A.Soloshenko, K.V.Stepanyantz},
Two-loop renormalization of $N=1$ super-symmetric electrodynamics,
regularized by higher derivatives, hep-th/0203118.

\bibitem{tmf2}
{\it A.A.Soloshenko, K.V.Stepanyantz}, Teor.Mat.Fiz. {\bf 134},
(2003), 429.

\bibitem{Siegel}
{\it W.Siegel}, Phys.Lett. {\bf 84 B}, (1979), 193.

\bibitem{Siegel2}
{\it W.Siegel}, Phys.Lett. {\bf 94B}, (1980), 37.

\bibitem{tHV}
{\it G.t'Hooft, M.Veltman}, Nucl.Phys. {\bf B44}, (1972), 189.

\bibitem{Nikolai}
{\it H.Nicolai, P.K.Townsend}, Phys.Lett. {\bf 93B}, (1980), 111.

\bibitem{Leveille}
{\it D.R.T.Jones, J.P.Leveille}, Nucl.Phys. {\bf B206}, (1982), 473.

\bibitem{West}
{\it P.West}, Introduction to supersymmetry and supergravity, World
Scientific, 1986.

\bibitem{MS}
{\it G.t'Hooft}, Nucl.Phys. {\bf B 61}, (1973), 455.

\bibitem{tmf1}
{\it A.A.Soloshenko, K.V.Stepanyantz}, Theor.Math.Phys., {\bf 131},
(2002), 631.

\bibitem{Bertlmann}
{\it R.Bertlmann}, Anomalies in quantum field theory, Clarendon press,
Oxford, 1996.

\bibitem{West_Paper}
{\it P.West}, Nucl.Phys. {\bf B 268}, (1986), 113.

\bibitem{Slavnov1}
{\it A.A.Slavnov}, Phys.Lett. {\bf B 518}, (2001), 195.

\bibitem{Slavnov2}
{\it A.A.Slavnov}, Theor.Math.Phys. {\bf 130}, (2002), 1.

\bibitem{SlavnovStepan1}
{\it A.A.Slavnov, K.V.Stepanyantz},
Universal invariant renormalization for super-symmetric theories,
hep-th/0208006.

\bibitem{DiffR}
{\it D.Z.Freedman, K.Johnson, J.I.Latorre}, Nucl.Phys. {\bf B371},
(1992), 353.

\bibitem{Mas}
{\it J.Mas, M.Perez-Victoria, C.Seijas}, JHEP, {\bf 0203}, (2002), 049,
(hep-th/0202082).

\end{thebibliography}
\end{document}